

\input harvmac
\input epsf

\def\p{^\prime}
\bigskip


%
%
\def\RF#1#2{\if*#1\ref#1{#2.}\else#1\fi}
\def\NRF#1#2{\if*#1\nref#1{#2.}\fi}
\def\refdef#1#2#3{\def#1{*}\def#2{#3}}
%
%
\def \ts{\thinspace}

\def \CMP{{\it Comm.\ts Math.\ts Phys.\ts }}
\def \NP{{\it Nucl.\ts Phys.\ts }}
\def \PL{{\it Phys.\ts Lett.\ts }}

\def \Tahoe{Proceedings of the XVIII
 International Conference on Differential Geometric Methods in Theoretical
 Physics: Physics and Geometry, Lake Tahoe, USA 2-8 July 1989}
\def \Tahoe{Proceedings of the NATO
 Conference on Differential Geometric Methods in Theoretical
 Physics, Lake Tahoe, USA 2-8 July 1989 (Plenum 1990)}
\def \Zm{Zamolodchikov}
\def \AZm{A.\ts B.\ts \Zm}
\def \AlZm{Al.\ts B.\ts \Zm}
\def\ped{P.\ts E.\ts Dorey}
\def\dur{H.\ts W.\ts Braden, E.\ts Corrigan, \ped\ and R.\ts Sasaki}

\refdef\rAFZa\AFZa{A.\ts E.\ts Arinshtein, V.\ts A.\ts Fateev and
 \AZm, {\it Quantum S-matrix of the $1+1$ dimensional
Toda chain}, \PL {\bf B87}
(1979) 389}

\refdef\rBSa\BSa{H. W. Braden and R. Sasaki, {\it The S-matrix
coupling dependence
for $ADE$ affine Toda field theory}, \PL {\bf B255} (1991) 343}

\refdef\rBSb\BSb{H.W. Braden and R. Sasaki, {\it Affine
Toda perturbation theory}, \hfill\break
\NP {\bf B379} (1992) 377}

\refdef\rBCDSa\BCDSa{\dur, {\it Extended Toda field
theory and exact S-matrices},
\PL {\bf B227} (1989) 411}

\refdef\rBCDSb\BCDSb{\dur, {\it Aspects of perturbed
 conformal field theory,
affine Toda field theory and exact S-matrices}, \Tahoe}

\refdef\rBCDSc\BCDSc{\dur, {\it Affine Toda field theory
and exact S-matrices},
\NP {\bf B338} (1990) 689}

\refdef\rBCDSe\BCDSe{\dur, {\it Multiple poles and other
 features of affine Toda
field theory}, \NP {\bf B356} (1991) 469}

\refdef\rCKKa\CKKa{H.S. Cho, I.G. Koh and J.D. Kim,
{\it Duality in the $d_4^{(3)}$
affine Toda theory}, KAIST preprint 1992}

\refdef\rCMa\CMa{P.\ts Christe and G.\ts Mussardo, {\it Integrable
sytems away from
 criticality: the Toda field theory and S-matrix of the
tri-critical Ising model}, {\it Nucl. Phys}.
 {\bf B330} (1990) 465}

\refdef\rCMb\CMb{P.\ts Christe and G.\ts Mussardo, {\it
Elastic S-matrices in $1+1$
 dimensions and Toda field theories},
 {\it Int.~J.~Mod.~Phys.}~{\bf A5} (1990) 4581}

\refdef\rCMc\CMc{J.\ts Cardy and G.\ts Mussardo, \\S-matrix
of the Yang-Lee
 edge singularity in two dimensions", \PL {\bf B225} (1989) 275}

\refdef\rCTa\CTa{S.\ts Coleman and H.\ts Thun, \CMP {\bf 61}
 (1978) 31}

\refdef\rDGPZa\DGPZa{G.W. Delius, M.T. Grisaru, S. Penati
and D. Zanon, {\it The
 exact S-matrices of affine Toda field theories based on
Lie superalgebras},
\PL {\bf B256} (1991) 164}

\refdef\rDGPZb\DGPZb{G.W. Delius, M.T. Grisaru, S. Penati and
D. Zanon, {\it Exact
S-matrices and perturbative calculations in affine Toda field
theories based on
 Lie superalgebras}, \NP {\bf B359} (1991) 125}

 \refdef\rDGZc\DGZc{G. W. Delius, M.T.   Grisaru, D.  Zanon,
{\it Exact S matrices
for non-simply laced affine Toda theories}, \NP {\bf B382}
(1992) 365}

\refdef\rDSa\DSa{V.\ts G.\ts Drinfel'd and V.\ts V.\ts Sokolov,
 {\it J. Sov. Math.} {\bf 30} (1984) 1975}

\refdef\rDDa\DDa{C.\ts Destri and H.\ts J.\ts de Vega, {\it
The exact S-matrix
of the affine $E_8$ Toda field theory}, {\it Phys. Lett.}
 {\bf B233} (1989) 336}

\refdef\rDa\Da{P.E. Dorey, {\it A remark on the coupling-dependence
in affine
Toda field theories}, CERN preprint April 1993}

\refdef\rDb\Db{\ped, Durham PhD thesis, unpublished}

\refdef\rDc\Dc{\ped, {\it Root systems and purely elastic
S-matrices},
\NP {\bf B358} (1991) 654}

\refdef\rDd\Dd{\ped, {\it Root systems and purely elastic
S-matrices II},
\NP {\bf B374} (1992) 741}

\refdef\rFZa\FZa{V.\ts A.\ts Fateev and \AZm, {\it Conformal
field theory
and purely elastic S-matrices}, {\it Int. J. Mod. Phys.} {\bf A5}
 (1990) 1025}

\refdef\rFOa\FOa{A. Fring and D. Olive, {\it The fusing
rule and scattering
matrix of affine Toda field theory}, \NP {\bf B379}
(1992) 429}

\refdef\rFLOa\FLOa{A. Fring, H.C. Liao and D. Olive,
{\it The mass spectrum
and coupling in affine Toda field theory},
\PL {\bf 266B} (1991) 82}

\refdef\rFKMa\FKMa{P. G. O. Freund, T. Klassen and E. Melzer,
{\it S-matrices
for  perturbations of certain conformal field theories},
{\it Phys. Lett.}
 {\bf B229} (1989) 243}

\refdef\rFb\Fb{M.\ts D.\ts Freeman, {\it On the mass
spectrum of affine Toda
field theory},\hfill\break
 \PL {\bf B261} (1991) 57}

\refdef\rHa\Ha{T.J. Hollowood, {\it Solitons in affine
Toda field theories},
\NP {\bf B384} (1992) 523}

\refdef\rKa\Ka{M.\ts Karowski, {\it On the bound state
problem in $1+1$
dimensional field theories}, \NP {\bf B153} (1979) 244}

\refdef\rKMa\KMa{T.\ts R.\ts Klassen and E.\ts Melzer,
{\it Purely elastic
scattering theories and their ultra-violet limits},
{\it Nucl. Phys.}
 {\bf B338} (1990) 485}

\refdef\rMa\Ma{P. Mansfield, {\it Nucl. Phys.} {\bf B222}
(1983) 419}

\refdef\rMMa\MMa{N.J.   MacKay, W.A.  McGhee, {\it  Affine
Toda solitons and
automorphisms of Dynkin diagrams}, Durham/Kyoto preprint, July 1992}

\refdef\rMOPa\MOPa{A. V. Mikhailov, M. A. Olshanetsky and
A. M. Perelomov,
{\it Two dimensional generalised Toda lattice}, {\it
 Comm. Math. Phys.} {\bf 79} (1981) 473}

\refdef\rOTa\OTa{D. I. Olive and N. Turok, {\it The symmetries
of Dynkin
diagrams and the reduction of Toda field equations}, {\it Nucl.
Phys.} {\bf B215} (1983)
 470}
\refdef\rOTc\OTc{D. I. Olive and N. Turok, {\it The Toda
lattice field
theory hierarchies and zero-curvature conditions in Kac-Moody
algebras},
{\it Nucl. Phys.} {\bf B265}
(1986) 469}

\refdef\rOTb\OTb{D. I. Olive and N. Turok, {\it Local conserved
densities
and zero-curvature conditions for Toda lattice field theories},
{\it Nucl. Phys.} {\bf B257} (1985) 277}

\refdef\rOTUa\OTUa{D. Olive, N. Turok, J.W.R. Underwood,
{\it Solitons and
the energy momentum tensor for affine Toda theory},
Swansea/Imperial
College preprint October 1992}

\refdef\rWa\Wa{G. Wilson, {\it The modified Lax and
two-dimensional Toda
lattice equations associated with simple Lie algebras}, {\it Ergod. Th.
and Dynam. Sys.} {\bf 1} (1981) 361}

\refdef\rZZa\ZZa{\AZm\  and \AlZm, {\it Factorised S-matrices
in 2 dimensions as
the exact solutions of certain relativistic quantum field
theory models},
{\it Ann. Phys.}
 {\bf 120} (1979) 253}

\refdef\rWKa\WKa{ H.G. Kausch and G.M.T. Watts, {\it
 Duality in quantum
Toda theory and $W$-algebras}, \NP {\bf B386} (1992) 166}

 \refdef\rWWa\WWa{ G.M.T. Watts,  R. A.  Weston, {\it $g_2^{(1)}$
affine
Toda field theory: A Numerical test of exact matrix results},
\PL {\bf B289} (1992) 61}


\rightline{DTP-93/19}
\rightline{YITP/U-93-09}
\medskip
\centerline{\bf ON A GENERALISED BOOTSTRAP PRINCIPLE}
\bigskip
\centerline{E. Corrigan$^1$, P.E. Dorey$^2$,   R. Sasaki$^{1,3}$}
\bigskip
\centerline{$^1$Department of Mathematical Sciences}
\centerline{University of Durham, Durham DH1 3LE, England}
\medskip
\centerline{$^2$Theory Division CERN}
\centerline{1211 Geneva 23, Switzerland}
\medskip
\centerline{$^3$Uji Research Center}
\centerline{Yukawa Institute for Theoretical Physics}
\centerline{Kyoto University, Uji 611, Japan}
\vskip 1in
\noindent{\bf Abstract}
\bigskip
The S-matrices for non-simply-laced affine Toda field
theories are considered
in the context of a generalised bootstrap principle.
The S-matrices, and in
particular their poles, depend on a parameter whose range
lies between the
Coxeter numbers of
dual pairs of the corresponding  non-simply-laced algebras.
It is proposed that
only odd order poles
in the physical strip with positive coefficients throughout
this range should
participate in the bootstrap. All other singularities have
an explanation
in principle
in terms of a generalised Coleman-Thun mechanism. Besides
the S-matrices
introduced by Delius, Grisaru and Zanon, the missing case
($f_4^{(1)},e_6^{(2)}$),
is also considered
and provides many interesting examples of pole generation.
\vfill\eject

\newsec{Introduction}

Affine Toda field
theory
\NRF\rAFZa
\AFZa\NRF\rMOPa{\MOPa\semi\OTa\semi\OTb\semi\OTc}\refs{\rAFZa
,\rMOPa}
is a theory of $r$ scalar fields in
two-dimensional Minkowski space-time, where $r$ is the rank
of a compact
semi-simple Lie algebra $g$.
The classical field theory is determined by the
lagrangian density
\eqn\ltoda{{\cal L}={1\over 2}
\partial_\mu\phi^a\partial^\mu\phi^a-V(\phi )}
where
\eqn\vtoda{V(\phi )={m^2\over
\beta^2}\sum_0^rn_ie^{\beta\alpha_i\cdot\phi}.}
In \vtoda , $m$ and $\beta$ are real, classically
 unimportant constants,
$\alpha_i\ i=1,\dots ,r$ are the simple roots of the
Lie algebra $g$,
and $\alpha_0=-\sum_1^rn_i\alpha_i$ is an integer
linear combination of the simple roots; it corresponds to
the extra spot
on an extended Dynkin-Kac diagram for $\hat g$. The coefficient
$n_0$
is taken
to be one. If the term $i=0$ is omitted from
\vtoda\ in
the lagrangian \ltoda , then the theory, both classically and after
quantisation is conformal; with the term $i=0$, the conformal symmetry
is broken but the theory remains classically integrable, in the sense
that there are infinitely many independent conserved charges in
involution\NRF\rWa{\Wa\semi\DSa}\refs{\rWa}. In the \lq real coupling'
Toda theory, the fields are
supposed to be real.
However, there have also been recent studies\NRF\rHa{\Ha\semi\MMa\semi\OTUa}
\refs{\rHa}
of the classical soliton
solutions to the equations of motion following from \ltoda ; for these,
the fields are complex. The discussion in this article will be restricted
to the real-coupling theories.

As quantum field theories,
the real-coupling affine Toda field theories fall into two classes.
There are those based on the simply-laced  root systems corresponding
to the diagrams for $a_n^{(1)},d_n^{(1)},e_n^{(1)}$, and the others
based on the non-simply laced root systems. These fall into dual pairs
(dual in this context meaning the replacement
$\alpha_i\rightarrow \check\alpha_i=2\alpha_i/\alpha_i^2$),
namely, $(b_n^{(1)},a_{2n-1}^{(2)}),\,
(c_n^{(1)},d_{n+1}^{(2)}),\, (g_2^{(1)},d_4^{(3)}),\,
(f_4^{(1)},e_6^{(2)})$, except for $a_{2n}^{(2)}$ which is self-dual.
The simply-laced root systems are also self-dual since $\alpha_i^2=2$.
Based on a bootstrap principle and a number of checks within
perturbation theory, it has proved possible to conjecture
\NRF\rBCDSa\BCDSa\NRF\rBCDSb\BCDSb\NRF\rBCDSc\BCDSc\NRF
\rCMa{\CMa\semi\CMb}\NRF\rDDa\DDa
\refs{\rAFZa ,\rBCDSa , \rBCDSb ,\rBCDSc ,\rCMa ,\rDDa}
the exact S-matrices for each affine Toda field theory
associated with a self-dual root system. These have many
interesting properties which have been extensively reviewed
elsewhere. In the context of this paper it is intended to
concentrate on just a couple of them.

The first important property concerns the bootstrap
\NRF\rKa\Ka\NRF\rZZa\ZZa\refs{\rKa, \rZZa}. The
S-matrix element $S_{ab}(\theta_a-\theta_b)$ corresponding
to the elastic scattering of a pair of particles $a,b$, is a
meromorphic function of the rapidity difference $\Theta
=\theta_a-\theta_b$. For real rapidity, $S_{ab}$ is a phase
given by a product of elementary blocks $\{ x\}$
defined by\refs{\rBCDSc}
\eqn\cblock{\{ x\} ={(x-1)(x+1)\over (x-1+B)(x+1-B)}\qquad
(x)={\sinh\left({\Theta \over 2}+{x\pi i\over 2h}\right)\over
\sinh\left({\Theta \over 2}-{x\pi i\over 2h}\right)}}
where $0\le B\le 2$, in the simply-laced cases, there is evidence
\NRF\rBSa{\BSa \semi\BSb}\refs{\rAFZa ,\rBCDSa , \rBCDSc ,\rCMa , \rBSa} for
\eqn\sdB{B(\beta )={1\over 2\pi}{\beta^2\over 1+\beta^2/4\pi},}
and $h=\sum_0^rn_i$ is the Coxeter number for the chosen
root system. In the self-dual theories, $S_{ab}$ has fixed
position poles and moving (coupling dependent) zeroes in the
physical strip ($0\le {\rm
Im}\Theta \le \pi$), and moving poles
and fixed position zeroes outside the physical strip.
The odd order poles,
with a coefficient equal to $i$ times a function of $B$
which is positive throughout the range of $B$,
$0\le B \le 2$ participate in the bootstrap. The
poles indicating a bound state \lq fusing' $ab\rightarrow c$
occur at precisely the rapidity necessary for energy-momentum
conservation given that the particles have mass ratios
identical with those derived from the classical lagrangian.
Moreover, such fusings occur if and only if there is a
corresponding three-point coupling between the three mass
eigenstates ($a,b,c$) in the classical lagrangian. The magnitude of a
three-point coupling is always classically proportional to the area of
the triangle whose sides have lengths equal to the masses
of the particles participating in the coupling. The
precise nature of the odd-order poles in a particular
scattering matrix element, and the existence of even order
fixed poles on the physical strip are explicable within
perturbation theory, in terms of Landau singularities.
Detailed checks have been made \NRF\rBCDSe\BCDSe\refs{\rBCDSe}
for second and third order
poles but not for the others (up to order twelve in the
theory associated with $e_8^{(1)}$). However, even in the
latter cases the origin of the poles is known in principle
in the sense that some Feynman diagrams in the perturbation
expansion have been identified which contribute to each of them.

Besides the bootstrap, the masses and the eigenvalues of the
conserved quantities\NRF\rKMa{\KMa}\refs{\rBCDSb ,\rKMa}
are known to be components of the
eigenvectors of the adjacency matrix of the simple roots
$\alpha_i,\ i=1\dots r$, a fact discovered classically for
the masses\NRF\rFb{\Fb\semi\FLOa}\refs{\rBCDSb ,\rBCDSc ,\rFb}
and which is preserved in the quantum field
theory. The possible couplings have been characterised
succinctly in \NRF\rDc\Dc\refs{\rDc}
where it is  noted that the particles of an
affine Toda field theory are each associated with an orbit
of a simple root under the action of the Coxeter element $w$ in
the Weyl group of the selected root lattice.  The S-matrix
itself may be expressed \NRF\rDd\Dd\refs{\rDc ,\rDd}
in several elegant ways using these
Coxeter orbits, but such expressions will not be needed
here.

Very few of these facts or formulae work in the quantum
theories based on root systems which are not self-dual and it
remains an
outstanding problem to find their generalisations. The classical data
for these types of affine Toda theory may be found in \rBCDSc ,
however, it
was clear in early renormalisation calculations that the masses of the
particles in these theories could not be in the same ratios as the
classical masses\refs{\rBCDSc ,\rCMa}. Hence the S-matrices would
have a different character from those of the self-dual theories.

Recently, Delius, Grisaru and Zanon \NRF\rDGZc\DGZc\refs{\rDGZc}
have made a number of fascinating conjectures
concerning the nature of the masses and S-matrices for
 the non-simply-laced
theories. One apparent  consequence of
their work is that each dual
pair of root lattices corresponds to a one parameter set of quantum
field theories with particle masses \lq floating' between the classical
masses of the partners in each pair (see also \NRF\rWKa{\WKa\semi\CKKa}
\refs{\rWKa}). This feature is compatible with
a generalised bootstrap principle which will be outlined below. The
candidate S-matrices were not obtained for all the pairs. The one which
in many ways is the most interesting (for the dual pair $f_4^{(1)},\,
e_6^{(2)}$), was omitted from \refs{\rDGZc} and  will be
described in detail in section(3). It displays a variety of properties
not shared by the other examples.

Besides the work of ref\refs{\rDGZc}, there is numerical evidence from
a simulation of the $g_2^{(1)}$ theory by Watts and Weston
\NRF\rWWa\WWa\refs{\rWWa} that
the masses in these theories based on dual pairs of root lattices do indeed
flow with the coupling $\beta$.

\newsec{Generalised bootstrap}

It was noted in the introduction that for the theories based on
self-dual lattices the odd order poles always participate in the
bootstrap and occur with coefficients whose sign does not change
as the coupling constant varies in the range $0\le\beta\le\infty$. The
direct channel poles have a positive coefficient (multiplied by $i$),
whereas the crossed channel poles have a negative coefficient. The
poles with this property will be referred to as positive
or negative definite, respectively.
Other
poles of even order have a real coefficient, do not participate in the
bootstrap and, together with higher order odd poles,
are explicable within standard perturbation theory
in terms of Landau singularities in Feynman
diagrams.

In the theories based on non-self-dual root systems, the masses must
float and therefore the S-matrices will have moving poles on the
physical strip. A criterion is needed to decide
on the basis of the S-matrix alone which moving poles should
participate in the bootstrap. Given the experience with the simply-laced
theories, it is reasonable to suppose the odd order poles are again the
relevant ones. However, as Delius et al. note, their conjectured
S-matrices contain poles of odd order on the physical strip which cannot
be consistently interpreted as bound states with a  mass as
indicated by the
position of the pole. Some comments on this will be made later. A
careful study of the coefficients of these poles reveals that they do
not share in all respects the characteristic behaviour noted in the
self-dual cases. Specifically, the
coefficients of these poles {\it change sign} at least once for some
value of the floating parameter as it floats between the two partners in
the dual pair. Such poles will be referred to as semi-positive. The other
odd order poles, which do participate in the
bootstrap, have coefficients which do not change sign over the floating
interval. As will be seen, this is a persistent feature in all possible
cases.

The theory associated with the pair $(g_2^{(1)},\, d_4^{(3)})$ will be
reviewed first to illustrate the ideas.

\newsec{The case $(g_2^{(1)},\, d_4^{(3)})$}

The floating masses for this theory are conjectured\refs{\rDGZc} to be
\eqn\gmasses{m_1=\sin\pi /H\qquad m_2=\sin 2\pi /H}
up to an overall factor $(2\surd 2 m)$ which is ignored. The parameter $H$
floats in the range $6\le H\le 12$, ie between the Coxeter numbers of
the partners in the pair. Each particle
couples to itself but there is also expected to be a fusing $1\, 1
\rightarrow 2$ in addition. The rapidity ($\Theta =2i\pi /H$)
at which the latter fusing
occurs also floats, since
\eqn\typea{m_2^2=\sin^22\pi /H=2m_1^2+2m_1^2\cosh\Theta =
2\sin^2(\pi /H)\ (1+\cos 2\pi /H).}
Each particle is self-conjugate.
Finally, it may be supposed that for $H=6$ or $12$, the S-matrix
elements are unity.

With these points in mind, the simplest choice\refs{\rDGZc} for $S_{11}$ is:
\eqn\gsll{S_{11}(\Theta )={(0)\ (2)\over (H/3\, -2)(4-H/3)}\
{(H/3)\ (2H/3)\over (H-4)(4)}\ {(H-2)\ (H)\over (2+2H/3)(4H/3\, -4)}}
where the bracket notation has been adjusted slightly and is now defined
by
\eqn\Hblock{(x)={\sinh\left({\Theta \over 2}+{x\pi i\over 2H}\right)\over
\sinh\left({\Theta \over 2}-{x\pi i\over 2H}\right)}.}
The terms in \gsll\ have been grouped together in order to facilitate
the inspection of the coefficients of the moving poles. In fact, for $H$
within the stated range, the poles at $\Theta =2\pi i/3$ and at $2\pi
i/H$ are positive definite direct channel poles throughout the
range of $H$. The self-coupling bootstrap for the pole at $2\pi /3$,
\eqn\llbootstrap{S_{11}(\Theta )= S_{11}(\Theta -i\pi /3)S_{11}(\Theta +
i\pi /3),}
is satisfied.

The fusing $1\, 1\rightarrow 2$ can be used to define the S-matrix
elements $S_{12}$ and $S_{22}$ via the bootstrap. In other words,
\eqn\gbootstrap{\eqalign{S_{12}(\Theta )&=S_{11}(\Theta -i\pi
/H)S_{11}(\Theta +i\pi /H)\cr
S_{22}(\Theta )&=S_{12}(\Theta -i\pi
/H)S_{12}(\Theta +i\pi /H).\cr}}
This gives:
\eqn\gslh{S_{12}(\Theta )= {(1)\ (2H/3\, -1)\over (H-5)(5-H/3)}\ {(H/3\,
+1)\ (H-1)\over (4H/3\, -5)(5)},         }
which has a physical pole with a positive residue at $\Theta =i\pi
(1-1/H)$, corresponding to the fusing $1\ 2\rightarrow 1$, together with
its crossed partner, and an extra pair of poles at
$\Theta =i\pi (2/3\, -1/H),\ i\pi (1/3\, +1/H)$. The first of the two
extra poles has a negative coefficient, the second positive for $H$ in
the range $6\le H\le 9$. On the other hand, for $9\le H\le 12$, the
coefficients of these
two poles have the opposite sign. These poles are therefore semi-positive.
This phenomenon occurs due to a
crossing over at $H=9$ of
the two factors $(H-5)$ and $(5)$ in the denominator of
\gslh\ with the two factors $(H/3\, +1)$ and $(2H/3\, -1)$, respectively
in the numerator.

The second relation of \gbootstrap\ gives:
\eqn\gshh{\eqalign{S_{22}(\Theta )={(0)\ (2H/3\, -2)\over
(H-6)(4-H/3)}\ &{(2)\ (2H/3)\over (H-4)(6-H/3)}\cr & {(H/3)\ (H-2)\over
(4H/3\, -6)\ (4)}\ {(H/3\, +2)\ (H)\over (4H/3\, -4)\ (6)}\cr}.}
This has a variety of simple poles on the physical strip but only one of
them, at $\Theta =2i\pi /3$,
has a positive residue throughout the range of $H$.
Another, its crossed partner,
has a negative residue throughout the range. The rest have
coefficients which change sign at least once (and sometimes twice) over
the interval.

Two of the semi-positive poles,
at rapidity value
$\Theta = i\pi (2/3 \, -1/H)$ in \gslh\ and at
$\Theta =i\pi (1-2/H)$ in \gshh , are
positive in the region near $H=12$ and reflect the existence
of an extra $221$ coupling in the $d_4^{(3)}$ classical
theory\refs{\rBCDSc} which
is absent in the $g_2^{(1)}$ theory. However, except at the limit
$H=12$ (where the S-matrix elements are unity), the position of
these poles do not coincide with the floating masses. Because these poles
are semi-positive, they should not participate in the bootstrap. They are
artifacts of the bootstrap which
can be explained in the same spirit as that invoked by
Coleman and Thun\NRF\rCTa\CTa\refs{\rCTa} to explain the double poles
of the Sine-Gordon
breather S-matrix elements. As is usual in Toda theory, the mechanism
is subtle and will be described in section(8).

The S-matrix elements \gsll ,\gslh , and \gshh\
look quite strange relative to the elegant
formulae in the self-dual cases in terms of the basic block of eq\cblock.
However, they may each be written in a similar fashion in terms of a
slightly more general block $\{ x\}_\nu$ defined as follows.
For this purpose, it is
convenient to set $H=6+3B$, and to
suppose $0\le B\le 2$. However, this parametrisation is not intended to
imply $B$ has the form suggested in \sdB . Then, define

\eqn\nublock{\{ x\}_\nu ={(x -\nu B -1)(x+\nu B +1)\over (x+ \nu
B+B-1)(x-\nu B-B+1)}.}
Clearly the old block \cblock\ corresponds to $\{ x\}_0$
when $H$ reverts to $h$. In terms of these new building blocks, the
S-matrix elements are
\eqn\gsall{\eqalign {S_{11}(\Theta )&=\{ 1\}_0\ \{ H/2\}_{1/2}\ \{
H-1\}_0\cr
S_{12}(\Theta )&=\{ H/3\}_1\ \{ 2H/3\}_1\cr
S_{22}(\Theta )&=\{ H/3\, -1\}_1\ \{ H/3\, +1\}_1\   \{ 2H/3\, -1\}_1\
\{ 2H/3\, +1\}_1.\cr}}
Unfortunately, although these expressions are quite compact, the
extended blocks do not behave so naturally as \cblock\ with respect to
the bootstrap.

Finally, it is worth remarking that the eigenvalues of the conserved
quantities, compatible with the bootstrap for the conserved charges are
\eqn\gcharges{q_1^s=\sin s\pi /H \qquad q_2^s=\sin 2s\pi /H}
where $s$, the spin of the conserved charges is equal to 1 or 5 mod 6.

\newsec{The case $(f_4^{(1)},\, e_6^{(2)})$}

It is proposed that the floating masses for the theory based on the pair
of non-simply laced root systems
$f_4^{(1)}$ and $e_6^{(2)}$
should be (up to an overall factor $\surd 3 m$),
\eqn\femasses{\eqalign{
&m_1=\sin \pi /H\, \sin 2\pi /H\p\cr
&m_2=\sin 3\pi /H\, \sin \pi /H\p\cr
&m_3=\sin 2\pi /H\, \sin 2\pi /H\p\cr
&m_4=\sin 3\pi /H\, \sin 2\pi /H\p\cr}}
where
\eqn\hprime{{1\over H}+{1\over H\p}={1\over 6}.}
It is not hard to check that in the range $12\le H\le 18$, and up to an
overall factor, the masses
float between the masses of the particles of the classical $f_4^{(1)}$
theory  and those of the $e_6^{(2)}$ theory, (as given in \refs{\rBCDSc}).
Again, it is sometimes convenient notationally
to set $H=12+3B$. However, as in the previous case,
it is not intended in this article to relate $B$ explicitly to
the coupling constant.

In this case, not all of the  couplings of either classical theory
will float throughout the range in $H$, and thus be able to participate
in the bootstrap. The relevant couplings that do float are:
\eqn\fecouplings{111,\ 112,\ 113,\ 123, \ 134, \ 222,\ 224, \ 333,\
444,}
with the corresponding fusing angles (in units of $\pi /H$)
\eqn\feangles{\eqalign{&U_{11}^2 = H/3\, +2,\ U_{12}^1 =5H/6\, -1,\cr
&U_{11}^3 =2, \ U_{13}^1 =H -1,\cr
&U_{22}^4 =H/3\, -2,\ U_{24}^2 =5H/6\, +1,\cr
&U_{12}^3 =H/2\, -1,\ U_{13}^2 =2H/3\, +1,\ U_{23}^1 =5H/6,\cr
&U_{13}^4 =3,\ U_{14}^3 =H-2,\  U_{34}^1 =H-1,\cr}}
and the self couplings correspond in the same units to the fusing angle
$2H/3$. Of course, these were actually discovered by checking the
consistency of the bootstrap in the sense indicated above. Here, the
S-matrix elements will be given but the detailed checks of the bootstrap
will be omitted.
Noting that all S-matrix elements are crossing symmetric, it is
convenient to set
\eqn\newblock{[x]_\nu =\{ x\}_\nu \{ H-x\}_\nu .}
In terms of this, the  S-matrix elements are given by
$$\eqalignno{
S_{11}(\Theta) &= [1]_0\,[H/3 + 1]_0 \cr
S_{12}(\Theta) &= [H/6+2]_0\,[H/2 + 2]_0 \cr
S_{13}(\Theta) &= [2]_0\,[H/3]_0\,[H/3 +2]_0 \cr
S_{14}(\Theta) &= [3]_0\,[H/3 -1]_0\,[H/3 + 1]_0\,[H/3 +3]_0 \cr
S_{22}(\Theta) &= [1]_0\,[H/3 -3]_0\,[H/3 + 1]_0\,[H/3 +3]_0 \cr
S_{23}(\Theta) &= [H/6 +1]_0\,[H/6 +3]_0\,[H/2+1]_0\,[H/2 + 3]_0 \cr
S_{24}(\Theta) &= [H/6]_0\, [H/6+2]_0\,[H/6+4]_0\,[H/2]_0\,[H/2+2]_0\,
[H/2 + 4]_0 \cr
S_{33}(\Theta) &= [1]_0\,[3]_0\,[H/3 - 1]_0\,[H/3+1]^2_0\,[H/3+3]_0 \cr
S_{34}(\Theta) &= [2]_0\,[4]_0\,[H/3 - 2]_0\,[H/3]_0^2\,[H/3+2]^2_0\,
[H/3+4]_0 \cr
S_{44}(\Theta) &= [1]_0\,[3]_0\,[5]_0\,[H/3-3]_0\,[H/3 - 1]_0^2\,[H/3+1]^3_0\,
[H/3+3]_0^2\,[H/3+5]_0. \cr
}$$
This expression is quite useful for checking the bootstrap properties
 but it conceals a
number of cancellations between zeroes and poles. An alternative,
in which all the
cancelling poles
and zeroes have been removed, is given by:
\eqn\sfematrixa{\eqalign{&S_{11}(\Theta )=[1]_0\, [H/3\, +1]_0  \cr
&S_{12}(\Theta )=[H/3]_{1/2}  \cr
&S_{13}(\Theta )=[2]_0\, [H/3]_0\, \{ H/2\}_{1/2}  \cr
&S_{14}(\Theta )=[H/6\,  +1]_{1/2}\, [H/2\, +1]_{1/2}  \cr
&S_{22}(\Theta )=[H/6\, -1]_{1/2}\, [H/2\, +1]_{1/2}  \cr
&S_{23}(\Theta )=[H/3\, -1]_{1/2}\, [H/3\, +1]_{1/2} \cr
&S_{24}(\Theta )=[H/3\, -2]_{1/2}\, [H/3\, ]_{1/2}\, [H/3\, +2]_{1/2}  \cr
&S_{33}(\Theta )=[1]_0\, [H/6\, +1]_{1/2}\, [H/3\, +1]_0\, [H/2\,
+11]_{1/2}  \cr
&S_{34}(\Theta )=[H/6]_{1/2}\, [H/6\, +2]_{1/2}\, [H/2\, +2]_{1/2}\,
[H/2]_{1/2}  \cr
&S_{44}(\Theta )=[H/6\, -1]_{1/2}\, [H/6\, +1]_{1/2}\,
[H/3]_{0}^\prime\,  [H/2\,
+1]_{1/2}^2\, . \cr }}
There is one curious set of terms, however, occuring in $S_{44}$ and
represented by
\eqn\atwoblock{[x]^{\prime}_0 =\{ x\}_0^\prime \{ H-x\}_0^\prime
\qquad
\{ x \}_0^\prime ={(x-2)\ (x+2)\over (x-2+2B)(x+2-2B)}.}
The latter type of
extended block occurs otherwise in only one other place,
in the S-matrices for $a^{(2)}_{2n}$.

The expressions given in \sfematrixa\ are very compact and conceal much
of the detailed information concerning the analytic structure. However,
there are no hidden cancellations that occur when these expressions are
expanded into their elementary ratios of hyperbolic sines. If these
expressions are examined carefully, there are no positive definite poles
other than those at rapidities given by the fusing angles \feangles .
However, some of the positive definite poles are not simple, those
corresponding to the self-couplings $333$ in $S_{33}$ and $444$ in
$S_{44}$ are cubic poles, while the rest are first order. The S-matrix
elements $S_{11},\ S_{12},\ S_{13}$ are straightforward, all poles are
either positive or negative definite except for the double poles in
$S_{13}$ at $\Theta =i(1/3 +1/H)\pi$ and $\Theta =i(2/3-1/H)\pi$, which
have a standard explanation in terms of the Coleman-Thun mechanism (see
below). The $S_{14}$ element has a positive definite pole at $\Theta
=i(1-2/H)\pi$ (and a negative definite pole at the crossed angle), a
pair of semi-definite poles at $\Theta =i (1/3+2/H)\pi$ and $\Theta
=i(2/3-2/H)\pi$, and a pair of double poles. $S_{22}$ is similar, with a
pair of semi-definite poles and four definite simple poles. $S_{23}$ has
a positive definite pole and its crossed partner, four semi-positive
poles and a double pole. $S_{24}$ is more intricate---there are six
semi-positive poles, two double poles and just one positive pole
corresponding to the coupling 242, and its crossed partner. $S_{33}$ has
four double poles and a positive definite cubic pole with its crossed
partner. $S_{34}$ has six semi-positive poles, two of them cubic, a pair
of double poles, and a positive definite simple pole with its crossed
partner, corresponding to the coupling 341. Finally, $S_{44}$ is the
most intricate despite having a single definite cubic pole with its
crossed partner; there is also a pair of semi-positive cubic poles, a
pair of semi-positive simple poles and a pair of double poles. Even the
positive definite cubic poles have that property as a consequence of a
miracle in which a pair of zeroes cross at $H=15$.

The eigenvalues of the conserved quantities
compatible with the bootstrap are given by
\eqn\feqvalues{\eqalign{
&q^s_1=\sin s\pi /H\, \sin 2s\pi /H\p\cr
&q^s_2=\sin 3s\pi /H\, \sin s\pi /H\p\cr
&q^s_3=\sin 2s\pi /H\, \sin 2s\pi /H\p\cr
&q^s_4=\sin 3s\pi /H\, \sin 2s\pi /H\p ,\cr}}
for $s= 1,5 $ mod 6. The possible spins of the conserved
quantities are the
exponents common to both $e_7$ and $e_6$.

\newsec{The pair $c_n^{(1)},\, d_{n+1}^{(2)}$}

For these cases, it was already proposed by Delius et al. that
the floating masses
ought, except for an overall factor, to be given by
\eqn\cmasses{m_a=\sin {a\pi\over H}\qquad a=1,2,\dots ,n}
where $H$ lies in the range $2n\le H\le 2n+2$. Also, in this case,
 with the
 same proviso as
before, $H=2n+B$.
The floating couplings are solely
of the type
\eqn\ccouplings{ab\rightarrow a+b\qquad\cases{U_{ab}^{a+b}=(a+b)\pi
/H&\cr
U_{a\ a+b}^b=(H-b)\pi /H&\cr U_{b\ a+b}^a=(H-a)\pi /H,&\cr}}
where $a+b\le n$, or their crossed partners.

The S-matrix elements given in\refs{\rDGZc} are
\eqn\csmatrixa{S_{ab}(\Theta )=\prod^{a+b-1}_{p=|a-b|+1\atop
{\rm step}\ 2}\, [p]_0.}
However, as in the previous cases, this notation conceals a
number of interesting
cancellations when $a+b\, >\, n$. For these elements, some of
the double poles are cancelled by zeroes to leave  semi-positive poles,
and some of the
simple poles disappear. For example, the pole which might occur in
\csmatrixa\ at
$\Theta =i(a+b)\pi /H$, for $a+b > n$ simply disappears. A series of
semi-positive simple poles, occuring
at $\Theta =i(2n+2-a-b)\pi /H,\ \dots\  i(a+b-2)\pi /H$ and
their partners,
should not be invited
to the bootstrap (even though there are classical couplings of this
 type in the lagrangian
for the $c_n^{(1)}$ affine Toda theory), but can be explained,
as will be
described in section(8), using a modified Coleman-Thun mechanism.
In terms of the modified blocks the
expressions for the S-matrix
elements with $a+b\, >\, n$ are
\eqn\csmatrix{S_{ab}(\Theta )=\prod^{2n-a-b-1}_{p=|a-b|+1\atop
{\rm step}\ 2}\, [p]_0\,
\prod^{n+1-a-b}_{p=a+b-n-1\atop \rm{step}\ 2}\, \{ H/2\, -p\}_{1/2}.}
The only positive poles occuring in these S-matrix elements are those
corresponding to
the couplings \ccouplings .

The eigenvalues of the conserved quantities compatible with the
bootstrap are
\eqn\cdqvalues{q^s_a=\sin {sa\pi\over H}}
where $s$ is any odd integer.

\newsec{The pair $b_n^{(1)}, a_{2n-1}^{(2)}$}

Here the spectrum contains one particle, labelled $n$, whose mass
is conveniently chosen not to float and the
others, labelled $1,2,\dots , n-1$, whose masses float, according to
Delius et al,
in the
following manner
\eqn\bmasses{ m_n=1\qquad m_a=2\sin {a\pi\over H},\quad a=1,2,\
dots ,n-1 ,}
with $2n-1\le H\le 2n$. In this case, $B=4n-2H$.

The floating couplings are the following:
\eqn\bcouplings{\eqalign{&nn\rightarrow a,\qquad U_{nn}^a=(H-2a)\pi
 /H\quad
U_{na}^n= (H/2+a)\pi /H
\quad a=1,2\dots ,n-1\cr
&ab\rightarrow a+b < n \qquad \cases{U_{ab}^{a+b}=(a+b)\pi /H
&\cr U_{a\ a+b}^b=(H-b)\pi /H
&\cr U_{b\ a+b}^a=(H-a)\pi /H.&\cr}\cr}}

The S-matrix elements are given by
\eqn\bsamatrix{\eqalign{&S_{nn}(\Theta )=\prod^{n-1}_{p=1-n\atop
{\rm step}\ 2}\{ H/2 -p\}_{-1/4}\cr
&S_{an}(\Theta )= \prod_{p=1\atop{\rm step\ 2}}^{2a-1}\,
\{{H\over 2}-a+p\}_0\cr
&S_{ab}(\Theta )= \prod^{a+b-1}_{p=|a-b|+1\atop {\rm step}\ 2}\,
[p]_0   \cr}}

In this case, in $S_{ab}$, there is no hidden cancellation
of zeroes and
poles. The S-matrices
$S_{nn}$ and $S_{an}$ have
positive definite simple poles, and their crossed partners.
The $S_{an}$
S-matrix elements also have
double poles
which are explained by the standard Coleman-Thun mechanism.
The S-matrix elements $S_{ab}$
have a series of double poles and a pair of semi-positive
poles at $i(a+b)\pi /H$ and $i(H-a-b)\pi /H$, when $a+b>n$.
These are also
explained in section(8). If $a+b<n$, these poles are
positive/negative
definite, respectively.

\newsec{The case $a_{2n}^{(2)}$}

The S-matrix corresponding to the theory based on the roots of
$a_{2n}^{(2)}$,
will be included here as one of the non-simply-laced cases, for
completeness.
It has a chequered history. The minimal S-matrix  for this theory
(ie just the terms without
dependence on $B$) is non-unitary, and was written down by
Freund, Klassen and Melzer \NRF\rFKMa\FKMa\refs{\rFKMa}; it
generalises the
Bullough-Dodd theory. Despite being a non-simply-laced theory,
the classical
and quantum mass ratios coincide \refs{\rCMa},
and the floating  does not occur. On the other hand, the
affine root system is
self-dual, and for that reason the floating would not be
expected given the new
insight furnished by \refs{\rDGZc}.

The S-matrix is\refs{\rCMa}:
\eqn\atwonsmatrix{S_{ab}(\Theta )=\prod_{|a-b|+2\atop{{\rm step}\
4}}^{a+b-2}\,
\{ p\}^\prime_0\{ 4n+2-p\}^\prime_0\qquad a,b=2,4,\dots , 2n}
using the block defined in \atwoblock\ with $H=h=4n+2$.
Actually, these S-matrix elements may
be thought of from the point of view of folding
$d_{2n+2}^{(1)}$ \refs{\rBCDSc}.
The spectrum of the $a_{2n}^{(2)}$ theory coincides with
the even-labelled
particles in the spectrum of the $d_{2n+2}^{(1)}$ theory.
The S-matrices of
the latter have a multiple pole structure
which is not explicable in terms of the truncated spectrum.
However, the
$a_{2n}^{(2)}$ S-matrix elements are obtained by simply deleting
the inexplicable
poles. This procedure leads to the above expression in terms of
\atwoblock , and does not
upset the bootstrap. Attempting to do the same in terms of the
foldings leading to the
other non-simply-laced theories simply does not work.

\newsec{Semi-positive and high order poles}

It has been noted in earlier sections that the S-matrices with
floating
couplings have a variety of pole
singularities. There are those of odd order whose coefficient
has a single
sign throughout the floating
range: these participate in the bootstrap, and indeed
consistently define
the floating couplings between
the particles of the theory. Others are double poles whose
origin lies in
the Coleman-Thun mechanism,
originally formulated to explain the double poles appearing
in the sine-Gordon S-matrix elements\refs{\rZZa}.
 Besides these, the S-matrices  have semi-positive simple
(and occasionally cubic) poles which do not
participate in the bootstrap and whose origin also relies
on the Coleman-Thun mechanism, with an
interesting extra subtlety not encountered previously.
One might say the prosaic explanation
for the double poles within the sine-Gordon theory
contains some poetry after all.

The basic mechanism relies on the existence of the diagram
in fig(1a),
in which the scattering particles $a,b$ each \lq defuse'
into the pairs $r,p$ and $r,q$ respectively,
with all particles simultaneously on shell. The circle in
the centre of the diagram represents the scattering
process in which the pair $p,q$ elastically scatter before
fusing with the particles $r$ to build the final
state of the elastic $a,b$ scattering process. The dual of
the on-shell scattering diagram is the figure (1b).
There, the triangles represent the couplings and have sides
whose lengths are the masses of the particles
$a,b,p,q $ and $r$, as indicated. The relative rapidity
of the pair $p,q$ as they scatter is $i$ times one of
the inner angles, labelled $\psi$ in the diagram. On the
other hand, the relative rapidity of the pair $a,b$
is $i$ times the angle labelled $\phi$. In terms of the
coupling angles, $\phi$ and $\psi$ are given by
\eqn\phipsi{\phi = \bar U^q_{rb}+\bar U^p_{ra}
\qquad \psi = U^b_{rq}+ U^a_{rp},}
where $\bar U=\pi -U$. In practice, such diagrams are
discovered
by inspecting the coupling triangles
and using them to provide a partial \lq tiling'
of the parallelogram
with sides $m_a,m_b$ and angles
$\phi$ and $\bar \phi$. Typically, the existence
of a diagram like fig(1a),
in which all lines may be
simultaneously on shell, explains the double pole.
The basic reasoning is
simple. On shell, according
to the Cutkosky rules, each internal propagating
particle  in the process
contributes a delta function
$\theta (p_0)\delta (p^2-m^2)$ and there are two loop
integrals, one for each
triangular loop  containing
the particles $p,q,r$. Overall therefore, the six
delta functions and two loop
integrals
combine to yield a double delta function which
translates to a double
pole in terms of the relative rapidity between
particles $a$ and $b$, ie $\Theta$.

The above argument represents the generic situation.
The $p,q$ scattering matrix
 element appearing in the
 middle of the diagram plays a r\^ ole only in the detailed
computations for
the coefficient of the double pole.
However, and this is the subtle part, the $p,q$ S-matrix
element may have a
zero at $\Theta_{pq}=i\psi$.
If that happens then the double pole will be reduced to a
simple pole. Moreover,
the coefficient of the
pole as a function of the floating parameter ($H$ or $B$)
 will be composed of
 two pieces.  First of all, there
are the four couplings occuring in pairs corresponding
 to $arp$ and $brq$.
These are computed from the
coefficients of positive poles in the S-matrix elements
$S_{rp}$ and $S_{rq}$,
corresponding to the fusings
$rp\rightarrow a$ and $rq\rightarrow b$, and, by
definition, these factors do not
change sign over the floating
interval. The other important factor enters as the
coefficient of the
zero in $S_{pq}$,
and this can change sign
over the floating interval. Indeed, this is precisely
what happens and
provides a mechanism for many of the
semi-positive poles. If $S_{pq}$ has a double zero, then
the potential
double pole will be removed completely.

However, this is not the whole story.  Occasionally, a
semi-positive
pole has a more complicated explanation
lying beyond the basic Coleman-Thun mechanism. It may
happen that there
is more than one rescattering in
the middle of an on-shell diagram. For example,
consider fig(2a) and its
dual diagram fig(2b).
 Here, each particle $a,b$ defuses into the pairs
$r,p$ and $r\p ,q$,
respectively,  and the pair $p,q$
then fuses to make $r^{\prime\prime}$, all particles
being on-shell.
The middle circle then represents a
three-to-three scattering with the three relative
rapidities $\Theta_{r,r\p},
\Theta_{r,r^{\prime\prime}},
\Theta_{r\p,r^{\prime\prime}}$ fixed by energy-momentum
conservation.
Since the S-matrix theory
is factorisable, the three-particle scattering S-matrix
is regarded as
the product of the three two-particle
S-matrices $S_{rr\p}(\Theta_{r,r\p})\, S_{r\p
r^{\prime\prime}}
(\Theta_{r\p ,
r^{\prime\prime}})\,
S_{rr^{\prime\prime}}(\Theta_{r,r^{\prime\prime}})$,
see fig(2c),
evaluated for the appropriate rapidity differences.
Specifically,
\eqn\psidash{\eqalign{\Theta_{ab}&=i(\bar U^r_{ap}+
\bar U^{r\p}_{bq}+
\bar U^{r^{\prime\prime}}_{pq})\cr
\Theta_{rr\p}&=i(\bar U_{pq}^{r^{\prime\prime}}+U_{pr}^a+
U_{qr\p}^b)\cr
\Theta_{r\p r^{\prime\prime}}&=i(U^{p}_{qr^{\prime\prime}}+
U^{b}_{qr\p})\cr
\Theta_{rr^{\prime\prime}}&=i(U^{q}_{pr^{\prime\prime}}+
U^{a}_{pr})\cr}.}
In the absence of any zeroes, the existence of such an
on-shell diagram
would imply a cubic pole.  To see this,
it is enough to remark the number of internal lines (13)
less loop
integrations (5) in fig(2c).
However, one or more of the three factors in the
three-to-three
S-matrix element may have a zero,
in which case the order of the pole is reduced.
That some semi-positive
poles are explained in this way
is a matter of inspection. Several examples are
included below. Generically,
diagrams such as fig(2a) will provide
cubic poles whose coefficient has a sign determined
by the behaviour of the
three inner S-matrix elements
over the range of $H$.  Occasionally,
the opposite is true, one or more of the rescattering
S-matrices may also
have a pole which indicates
that one or more of the three inner parts of fig(2b)
(the parallelograms
marked with the dashed lines),
may also be fully or partially tiled. In those cases,
there may be an inner fusing (as for example in
figs(3e,3f)) or,
the Coleman-Thun
mechanism may be repeated (as for example in fig(3g)).
Note, there are
actually two
ways of  tiling the central part of fig(2b) using
the parallelograms.
However, they correspond to identical products of
S-matrix elements.

As a first example, consider the $S_{12}$ matrix element
in the
($g_2^{(1)},d_4^{(3)}$) theory, eq\gslh .
This has a semi-positive pole at $(1+H/3)i\pi /H$.
On the other
hand, there is a diagram of type fig(1a)
in which $p,q,r$ are each particle 1, and the angle
$\psi =(2+2H/3)\pi /H$.
At the latter angle, the matrix
element $S_{11}$, eqn\gsll , clearly has a simple zero,
which on inspection
is seen to change sign at $H=9$.
The $S_{22}$ matrix element exhibits both types of
behaviour. Consider the
semi-positive pole at $2i\pi /H$.
There is a diagram of type fig(1a) and again the internal
particles
are all particle 1; this time, $\psi =4\pi /H$ and an
inspection of $S_{11}$
reveals a zero at this angle. Again it
has a coefficient which changes sign over the floating range.
On the other hand,
there is also a
semi-positive pole at $\psi =(2H/3 -2)\pi /H$, and this
time there is a process
corresponding to the diagram fig(2a).
All the internal lines represent particle 1, and the inner
angles are given by
$$\Theta_{rr^{\prime\prime}}=\Theta_{r^\prime
r^{\prime\prime}}=i(2H/3 +2)\pi
/H\qquad \Theta_{rr^{\prime}}
=i(2H/3 -4)\pi /H,$$
and thus it is clear the three-to-three scattering
matrix has a double zero,
since two of its factors vanish.
This time the change of sign cannot come from the coefficients
of these zeroes
individually since only the
square of the coefficient enters. However, the third factor
$S_{rr^\prime}(i(2H/3 -4)\pi /H)=
S_{11}(i(2H/3 -4)\pi /H)$
does change sign over the interval (twice  in fact). Apart from
these singularities
(and their crossed partners),
the poles in the ($g_2^{(1)},d_4^{(3)}$) S-matrix elements are
positive or negative definite.

It has already been noted that the situation in the
($f_4^{(1)},e_6^{(2)}$)
 S-matrix
is much more involved.
The element $S_{13}$ has a standard double pole at
$i(2H/3 -1)\pi /H$. $S_{14}$
has a standard double
pole at $2i\pi /3$, and a semi-positive pole at
$i(H/3 +2)\pi /H$ which
is explained
by the above mechanism
(with $p,q =1,3$ in fig (1a)) since
$S_{13}(i(2/3+3/H)\pi )=0$. The
semi-positive pole in
$S_{22}$ at $i(H/3 +2)\pi /H$ is explained similarly
(with $p,q =1,1$
in fig(1a)),
since $S_{11}(i(2/3+4/H)\pi )=0$.
$S_{23}$ has a standard double pole at $i\pi /2$ and a pair of
semi-positive poles at
$i(H/6+2)\pi /H$ and
$i(H/2-2)\pi /H$ explained (with $p,q =1,1$and $p,q =1,3$,
respectively in fig(1a)), since
$S_{11}(i(1/3+4/H)\pi )=0$ and $S_{13}(i(2/3+3/H)\pi )=0$. $S_{24}$
has standard double
poles at $i(H/2\pm 1)\pi /H$, and semi-positive poles at
 $i(H/6+1)\pi /H$,
$i(H/6+3)\pi /H$
and
$i(H/2+3)\pi /H$ which are explained (with
$p,q =1,1, \ p,q =1,3$ and $p,q=3,3$, respectively
in fig(1a)),
since
$S_{11}(i(2/3+4/H)\pi )=0$, $S_{13}(i(1/3+5/H)\pi )=0$
and $S_{33}
(i(2/3+4/H)\pi )=0$. $S_{33}$
has a pair of standard double poles at $2i\pi /H$ and
$i(H/3+2)\pi /H$,
and a standard
(ie positive coefficient)
cubic pole at $2i\pi /3$. The latter cubic pole is
interesting for
another reason. Alone
among all the floating
couplings in these theories, the $333$ coupling in
($f_4^{(1)},e_6^{(2)}$)
has a subtiling,
displayed in fig(3a).
Because of this, there are several processes contributing
to the cubic pole,
among them the
pair of vertex
corrections in fig(3b). This sub-tiling of the $333$ mass
triangle also
complicates the issue
in a discussion of
the poles in the $S_{34}$ and $S_{44}$ matrix elements.
Specifically,
the $S_{34}$ element has
a standard
double pole at  $i(1/3-1/H)\pi $, and its crossed partner; a
semi-positive pole at $i3\pi /H$
explained
(with $p,q=1,3$ in fig(1a)), since $S_{13}( i5\pi /H)=0$; a
semi-positive pole at
$i(1/3+3/H)\pi$,
explained by the mechanism corresponding to fig(2b),with
$r,r^\prime ,
r^{\prime\prime}=3,1,1$,
since
$S_{13}(i(2/3+3/H)\pi )=0=S_{13}(i(2/3-5/H)\pi )$; and a
semi-positive cubic pole
at $i(H/3+1)\pi /H$ for which
there are several contributions exemplified by figs(3c,3d).
The diagram fig(3c)
yields a cubic pole in a
straightforward manner; the contribution corresponding to
fig(3d) relies on the
retiling of the $333$ coupling,
and a zero at $i(2/3+3/H)\pi$ in the $S_{13}$ matrix
element to reduce
the expected
fourth order pole
down to cubic. The coefficient of the zero changes sign over the
floating interval.
Finally, $S_{44}$
has a standard double pole at $i(H/3-2)\pi /H$ explained
(with $p,q=2,2$ in fig(1a));
a semi-positive pole
at $i2\pi /H$ explained (with $p,q=1,1$ in fig(1a)),
 since $S_{11}(6i\pi /H)=0$;
and cubic poles at $2i\pi /3$
and $i(2H/3-2)\pi /H$, the latter being semi-positive and the
former corresponding
to the coupling $444$.
Fig(3e) again makes use of retiling the $333$ coupling triangle
to represent a
process similar to that of
fig(2a) but including a couple of vertex corrections. The
expected fifth order
pole at $i(2H/3-2)\pi /H$
is reduced to a cubic pole
by a pair of zeroes (at $i(2/3+3/H)\pi $) in the two $S_{13}$
S-matrix elements
contributing to the
three-to-three scattering in the middle of the diagram. On
the other hand, fig(3f)
allows a cubic pole
at $2i\pi /3$, since the expected fourth order pole is reduced
to cubic by a zero
(at $i(2/3+4/H)\pi$) in the $S_{11}$
rescattering process inside the diagram.  It is also possible
 for a potential
singularity to be removed completely by the zeroes in the
rescattering S-matrices.
Two cases have been encountered. The first is in $S_{44}$
at $\Theta= i4\pi/H$.
The standard Coleman-Thun mechanism (fig(1a) with $r=1$,
$p=q=3$) fails to
produce any singularity here because $S_{33}$ has a double
zero at
$\psi=i6\pi/H$. The second, and more interesting, case
is again in $S_{44}$,
at $\Theta=i(H/3 + 4)\pi/H$, which has fig(2a) with
$r=r^\prime=3$,
$r^{\prime\prime}=p=q=1$. The potential cubic pole is removed
completely by the
three zeroes $S_{13}(i(2H/3+3)\pi/H)=0$ (twice) and
$S_{33}(i(H/3 +6)\pi/H)=0$.

Next, consider the pair ($c_n^{(1)},d_{n+1}^{(2)}$).
All the additional
singularities of $S_{ab}$ in \csmatrixa\ or \csmatrix ,
the double poles and
semi-positive simple poles, can be explained by one type
 of diagram, fig(1a) with
$r=k$, $p=a-k$, $q=b-k$, $\phi=(a+b-2k)\pi/H$ and
 $\psi=(a+b)\pi/H$ for
 various possible $k$'s.
When $a+b \le n$, they give the standard double
 poles for the entire range
of $k$, namely from $k=1,2,\dots , {\rm min}\ (a,b) -1$.
While, for $a+b > n$, the rescattering
S-matrix in the middle of the Coleman-Thun mechanism,
 $S_{a-k,\,b-k}$ has a zero
at $i\psi= i(a+b)\pi/H$ for $k=1,2,\dots ,a+b-n-1$.
In this range of $k$,
the processes corresponding to the diagrams provide a
series of
semi-positive simple poles located at
$i(2n -a-b+2)\pi/H ,\dots , i(a+b-2)\pi/H$. For
$k=a+b-n, \dots ,
{\rm min}\ (a,b) -1$, the element $S_{a-k,\, b-k}$
is non-vanishing at
$i\psi= i(a+b)\pi/H$ and a series of standard double poles at
$i(a-b+2)\pi/H, \dots , i(2n-a-b)\pi/H$ is generated.

The singularity structure of the S-matrices for the pair
($b_n^{(1)},a_{2n-1}^{(2)}$)
is the simplest. $S_{an}$ has a series of double poles at
$i(H/2-a+2k)\pi /H,\ k=1,2,\dots ,a-1$.
They are explained by two types of diagrams, the standard
Coleman-Thun crossed
box together with an uncrossed box.
$S_{ab}$ has a series of double poles at $i(|a-b| +2)\pi/H,
\dots ,i(a+b-2)\pi/H$,
which can be explained by the standard Coleman-Thun
 mechanism as in the pair ($c_n^{(1)},d_{n+1}^{(2)}$).
The only semi-positive simple poles are
at $i(a+b)\pi/H$ with its crossed partner for $a+b >n$.
It corresponds to $p=q=r=n$ and $\psi=(2 -2(a+b)/H)\pi$.
The S-matrix, $S_{nn}$, in the middle of fig(1a) has a simple zero at
$i\psi=i(2 -2(a+b)/H)\pi$, which reduces the singularity of
 $S_{ab}$ at
$i\phi= i(a+b)\pi/H$ to a simple pole. This is the pole whose
\lq shifted position' is discussed in detail from a different
point of view in \refs{\rDGZc}.

A detailed checking of the coefficients of
these poles will not be attempted here.
Nevertheless, for each singularity,
there is at least one identified
process that can produce it, and sometimes several.
There is no situation in which a semi-positive simple or cubic
pole fails
to have an explanation in these terms, at least in principle.

\newsec{Discussion}

The scattering matrices for the dual pairs are reminiscent of
the sine-Gordon breather S-matrices. There, because the spectrum
of the theory
is not perturbative, it was never expected to find a
 perturbative understanding
of the multiple pole singularities. Here, the mass-spectrum
floats with a parameter,
$H$ or $B$, interpolating the mass sets of the
two partners in the dual pair,
and is also expected to be non-perturbative. An infinite order of
perturbation theory would be needed to see the floating phenomenon
over its whole range. The self-dual affine
Toda theories are very special, the masses in the full
theory occur in the same
ratios as in the classical lagrangian,
there is no floating, and therefore the existence of the
multiple poles in the S-matrix can be inferred from the
perturbation theory relatively easily.
In those cases, it is the coefficients of the poles that
can be computed to finite order
only in the coupling constant, $\beta$.

Despite these differences, there appears to be a
general statement that
can be made
concerning the bootstrap. Namely, the existence
 of a genuine bound-state
fusing is signalled
by the existence of a pole of odd order, whose coefficient
is positive
throughout the
range of a parameter, which  interpolates
two theories with unit S-matrix. This statement applies
equally well to
both types of theory. Poles of odd order whose coefficients change
sign within this interval are not supposed to take part in the
bootstrap, which is
consistent without including them. Moreover, all such
singularities have an
explanation in terms
of a generalised Coleman-Thun mechanism in which the
 floating zeroes of the S-matrix
elements play a crucial r\^ ole. The fact that zeroes
 play a part emphasises the
non-perturbative nature of the mechanism since a zero
in an S-matrix element
is not easy to see in perturbation theory. It was perhaps
 fortunate that for the
simply-laced  theories these special mechanisms are not
apparently necessary
since all odd order poles in those cases have a coefficient
 of a definite sign
throughout the range of the coupling constant.

\bigskip
\noindent{\bf Acknowledgements}
\bigskip
All of us are grateful for the opportunity to spend some time at the
Isaac Newton Institute
for Mathematical Sciences, where this study was begun, and to
 G\' erard Watts for discussions. One of us (PED) is supported
by a European Community Fellowship, another  (RS) thanks
the Japan Society
for Promotion of Science for a one year Visiting Fellowship to the
University of Durham, and also the Department of Mathematical Sciences
for its hospitality.

\vfill\eject
\epsfverbosetrue

\listrefs
\vfill\eject
\nopagenumbers
\def\epsfsize#1#2{\hsize}
\epsffile{cds1.eps}
\vfill\eject
\epsffile{cds2.eps}
\vfill\eject
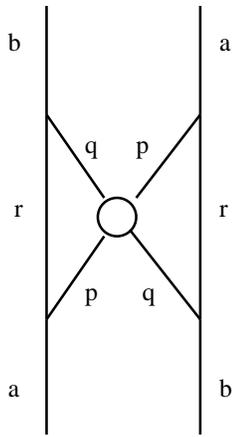
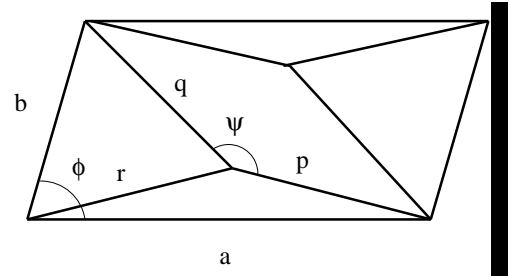
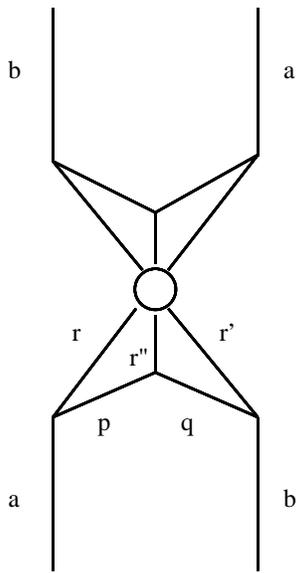
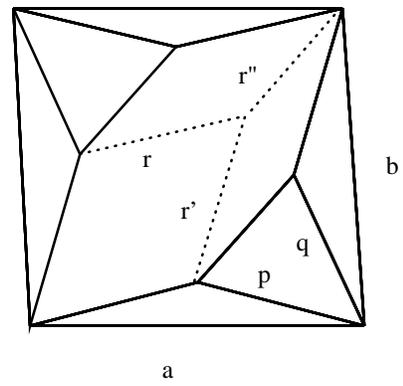
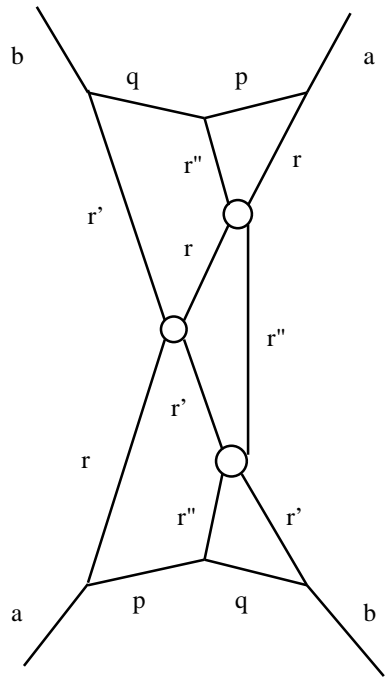
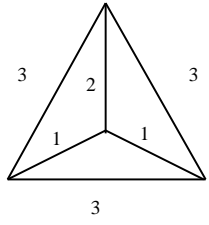
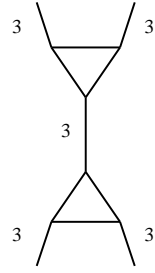
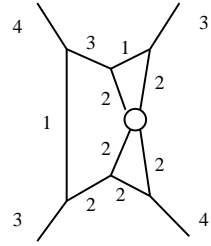
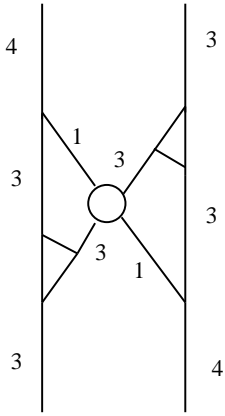
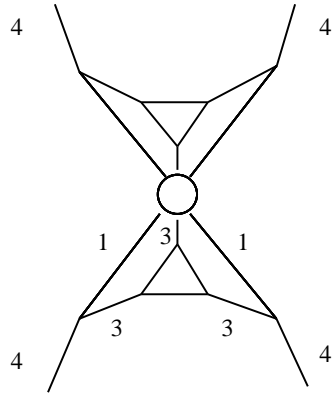
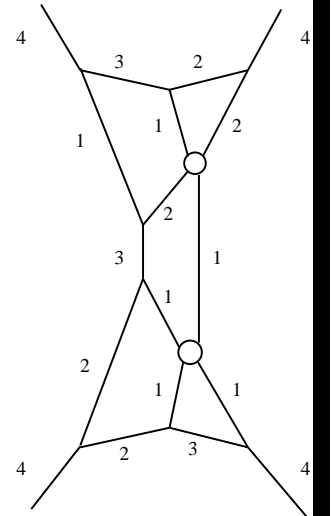
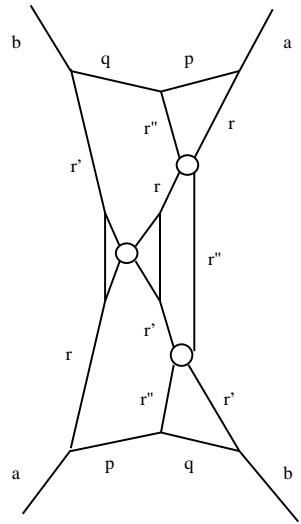
\end